\documentclass[aps,twocolumn,showpacs,preprintnumbers,nofootinbib,float,longbibliography,superscriptaddress]{revtex4-2}

\usepackage{epsfig}
\usepackage[dvipsnames]{xcolor}
\usepackage{float,amsmath,amssymb}
\usepackage{graphicx}
\usepackage{url}
\usepackage{bbold}
\usepackage[utf8]{inputenc}
\usepackage{physics}

\usepackage{subfig}

\newcommand{\by}{\mathbf{y}}

\newcommand{\xpom}{x_\mathbb{P}}
\newcommand{\gev}{\mathrm{GeV}}

\newcommand{\lqcd}{\Lambda_{\mathrm{QCD}}}
\newcommand{\as}{{\alpha_{\mathrm{s}}}}
\newcommand{\rt}{{\mathbf{r}_\perp}}

\newcommand{\xt}{{\mathbf{x}_\perp}}
\newcommand{\kt}{{\mathbf{k}_\perp}}
\newcommand{\yt}{{\mathbf{y}_\perp}}
\newcommand{\bt}{{\mathbf{b}_\perp}}
\newcommand{\bti}{{\mathbf{b}_{\perp,i}}}
\newcommand{\Deltat}{{\boldsymbol{\Delta}_\perp}}

\newcommand{\nc}{{N_\mathrm{c}}}
\newcommand{\jpsi}{$\mathrm{J}/\psi$ }
\newcommand{\jpsim}{\mathrm{J}/\psi}
\newcommand{\btheta}{{\boldsymbol{\theta}}}

\newcommand{\nf}{N_\mathrm{f}}
\newcommand{\Kwf}{K}
\newcommand{\mjimwlk}{m_\mathrm{JIMWLK}}
\usepackage[breaklinks,colorlinks,citecolor=citcolor,urlcolor=blue,linkcolor=lcolor]{hyperref}
\definecolor{lcolor}{rgb}{0.5,0,0}
\definecolor{citcolor}{rgb}{0,0.3,0.0}

\begin{document}

\title{Global Bayesian Analysis of $\mathrm{J}/\psi$ Photoproduction on Proton and Lead Targets}

\author{Heikki M\"antysaari}
\affiliation{Department of Physics, University of Jyv\"askyl\"a, P.O. Box 35, 40014 University of Jyv\"askyl\"a, Finland}
\affiliation{Helsinki Institute of Physics, P.O. Box 64, 00014 University of Helsinki, Finland}

\author{Hendrik Roch}
\affiliation{Department of Physics and Astronomy, Wayne State University, Detroit, Michigan 48201, USA}

\author{Farid Salazar}
\affiliation{Department of Physics, Temple University, Philadelphia, Pennsylvania 19122, USA}
\affiliation{RIKEN-BNL Research Center, Brookhaven National Laboratory, Upton, New York 11973, USA}
\affiliation{Physics Department, Brookhaven National Laboratory, Upton, New York 11973, USA}
\affiliation{Institute for Nuclear Theory, University of Washington, Seattle, Washington 98195, USA}

\author{Bj\"orn Schenke}
\affiliation{Physics Department, Brookhaven National Laboratory, Upton, New York 11973, USA}

\author{Chun Shen}
\affiliation{Department of Physics and Astronomy, Wayne State University, Detroit, Michigan 48201, USA}

\author{Wenbin Zhao}
\affiliation{Nuclear Science Division, Lawrence Berkeley National Laboratory, Berkeley,
California 94720, USA}
\affiliation{Physics Department, University of California, Berkeley, California 94720, USA}

\begin{abstract}
    We perform a global Bayesian analysis of diffractive $\mathrm{J}/\psi$ production in $\gamma+p$ and $\gamma+\mathrm{Pb}$ collisions using a color glass condensate (CGC) based calculation framework. As past calculations have shown that CGC-based models typically overpredict the $\mathrm{J}/\psi$ production in $\gamma+\mathrm{Pb}$ collisions at high center of mass energy, we address the question of whether it is possible to describe coherent and incoherent diffractive $\mathrm{J}/\psi$ data from $\gamma+p$ collisions at HERA and the LHC, and from $\gamma+\mathrm{Pb}$ collisions at the LHC simultaneously. Our results indicate that a simultaneous description of $\gamma+p$ and $\gamma+\mathrm{Pb}$ data is challenging, with results improving when an overall $K$-factor -- scaling $\gamma+p$ and $\gamma+\mathrm{Pb}$ cross sections to absorb model uncertainties -- is introduced.
\end{abstract}

\maketitle

\section{Introduction}
The quest to identify unambiguous experimental signatures of gluon saturation in high–energy quantum chromodynamics (QCD) has driven intense experimental and theoretical activity over the past two decades. 
At small parton momentum fractions $x$, the rapid growth of the gluon density predicted by linear evolution equations must eventually be tamed by non-linear recombination effects~\cite{Gribov:1983ivg,Mueller:1985wy}, giving rise to a semi-classical, high-occupancy regime known as the Color Glass Condensate (CGC)~\cite{McLerran:1993ni,McLerran:1993ka,Iancu:2003xm,Morreale:2021pnn,Garcia-Montero:2025hys}.

Diffractive vector meson production is a rather clean probe with a simple final state. 
Besides its cross section's sensitivity to the square of the target's gluon distribution (at leading order)~\cite{Ryskin:1992ui}, which originates from the requirement of a color-neutral exchange, it is sensitive to the spatial structure of the target~\cite{Mantysaari:2020axf}. 
Comparing this process between proton and heavy nuclear targets provides insight into the nuclear suppression mechanism. 
This mechanism could be nuclear shadowing~\cite{Ryskin:1992ui,Frankfurt:2003qy,Guzey:2013qza,Eskola:2022vpi,Guzey:2024gff} or gluon saturation~\cite{Lappi:2013am,SampaiodosSantos:2014puz,Cepila:2017nef,Luszczak:2019vdc,Bendova:2020hbb,Mantysaari:2022sux,Mantysaari:2023xcu}. 
For a comparison of different scenarios and models, see e.g. Ref.~\cite{Schenke:2024gnj}.

CGC-based calculations were used to describe the coherent and incoherent diffractive $\mathrm{J}/\psi$ production in $\gamma+p$ scattering, the latter being sensitive to fluctuations of the scattering amplitude~\cite{Mantysaari:2016ykx,Mantysaari:2017cni}, including those of geometric origin~\cite{Mantysaari:2018zdd,Mantysaari:2023qsq}. 
The model parameters in these calculations determine the evolution speed with the energy scale, the average and fluctuations of the geometry, as well as gluon density normalization and regulators of infrared physics related to confinement. 
A good fit to HERA data could be found, and the importance of a fluctuating proton substructure~\cite{Mantysaari:2018zdd} was established. 

The same framework has been used to make predictions for nuclear targets, as realized in ultraperipheral heavy-ion collisions, where one nucleus acts as a source of quasi-real photons which then interact with the other nucleus~\cite{Klein:2019qfb}. 
Experiments can extract the photon-nucleus center-of-mass energy, $W$, dependent $\jpsim$ production~\cite{Guzey:2013jaa,ALICE:2023jgu,CMS:2023snh} cross section of the $\gamma+\mathrm{Pb}$ system, and calculations constrained by HERA data generally predict a more rapid increase with $W$ than is observed experimentally~\cite{Lappi:2013am,Mantysaari:2022sux,Mantysaari:2023xcu,Mantysaari:2024zxq}. 
This means that the CGC calculations typically underestimate the degree of nuclear suppression observed in the experimental data.

This tension has not been resolved to date, motivating the current work.\footnote{Recently, it has been argued in Ref.~\cite{Luszczak:2024kgi} that including a subset of next-to-leading order (NLO) corrections may resolve this issue, but this conclusion is not confirmed by a full NLO calculation~\cite{Lappi:2021oag}.} 
Here, we set out to perform a combined Bayesian analysis of both $\gamma+p$ and $\gamma+\mathrm{Pb}$ diffractive $\jpsim$ production data, both coherent and incoherent, to determine whether one can find any set of parameter values that allows for the simultaneous description of both systems. 

We will find that within this general framework achieving a satisfactory simultaneous description of the $\gamma+p$ and $\gamma+\mathrm{Pb}$ data proves challenging, thereby motivating the introduction of a scale factor $K$ to uniformly rescale the cross sections for both proton and heavy‐ion targets.
The $K$-factor is also motivated by the existence of uncertainties stemming, for example, from the not well-constrained vector meson wave function and/or from higher-order effects in the perturbative expansion, which it can absorb. It is determined along with the other parameters in the Bayesian analysis. 
Its preferred value will lead us to conclude that the tension between $\gamma+p$ and $\gamma+\mathrm{Pb}$ data within our model cannot be easily removed by such scale uncertainty. 
We also demonstrate that several other extensions to the non-perturbative description of the nucleon structure at the initial energy scale do not remove the tension between the two collision systems and are not preferred by the experimental data.

This paper is organized as follows. 
In Sec.~\ref{sec:setup}, we review the theoretical framework used to calculate exclusive vector meson production and detail the various extensions introduced in this work.
Section~\ref{sec:Bayesian_inference} describes the Bayesian inference methodology, emulator training strategy, and validation procedures. 
We present our main results in Sec.~\ref{sec:results}, including posterior distributions for the model parameters and comparisons to experimental data.
Finally, we summarize our findings and discuss future directions in Sec.~\ref{sec:conclusions}.

\section{Exclusive vector meson production at high energy}
\label{sec:setup}
We calculate exclusive \jpsi photoproduction following Refs.~\cite{Mantysaari:2022sux,Mantysaari:2023xcu}, and briefly review the model in this section. 
We also introduce various extensions to the non-perturbative aspects of this model in order to determine if a more flexible initial state model can resolve the tension between the $\gamma+p$ and $\gamma+\mathrm{Pb}$ data. 

The cross section for coherent vector meson production differential in Mandelstam $t$ reads
\begin{equation}
\label{eq:coh}
    \frac{\dd\sigma^{\gamma^\ast + A \to \jpsim + A}}{\dd t} = \frac{\Kwf}{16\pi} \left| \left\langle \mathcal{A}^{\gamma^*+A \to V + A} \right\rangle_\Omega  \right|^2.
\end{equation}
Similarly, the incoherent cross section is given by
\begin{multline}
\label{eq:incoh}
    \frac{\dd\sigma^{\gamma^\ast + A \to \jpsim + A^*}}{\dd t} = \frac{\Kwf}{16\pi} \left(  
      \left\langle\left| \mathcal{A}^{\gamma^*+A \to V + A^\ast}  \right|^2\right\rangle_\Omega \right. \\
      -
    \left. \left| \left\langle \mathcal{A}^{\gamma^*+A \to V + A^\ast} \right\rangle_\Omega  \right|^2 \right).
\end{multline}
Here, $\langle \mathcal{O}\rangle_\Omega$ refers to the average over target configurations $\Omega$~\cite{Good:1960ba,Caldwell:2010zza,Mantysaari:2020axf}. 
In Eqs.~\eqref{eq:coh} and~\eqref{eq:incoh} we have introduced a constant scaling factor $\Kwf$. 
It can be interpreted, for example, as parametrizing uncertainty in the non-perturbative vector meson wave function discussed below (different wave function models typically result in exclusive cross sections that differ by an approximately constant factor~\cite{Lappi:2013am}).\footnote{We note that the vector meson wave function is constrained by the decay width data and by the normalization condition, such that the $\Kwf$ factor should not be interpreted as a multiplicative factor in the wave function itself, but as a means to describe the effect of changing the wave function on the cross section.}
On the other hand, this factor can also be interpreted as capturing potentially missing higher-order effects (see  Ref.~\cite{Mantysaari:2021ryb,Mantysaari:2022kdm} for a calculation of the cross section at NLO accuracy).

The scattering amplitude for diffractive vector meson $V$ production is given by~\cite{Kowalski:2006hc,Hatta:2017cte}
\begin{multline}
\label{eq:jpsi_amp}
    \mathcal{A}^{\gamma^*+p \to V + p} = 2i\int \dd[2]{\rt} \dd[2]{\bt}  \frac{\dd{z}}{4\pi} e^{-i \left[\bt - \left(\frac{1}{2}-z\right)\rt\right]\cdot \Deltat} \\
    \times [\Psi_V^* \Psi_\gamma](Q^2,\rt,z) N_\Omega(\rt,\bt,\xpom).
\end{multline}
Here $\rt$ describes the transverse size and orientation of the $q\bar q$ dipole, $\bt$ points from the center of the target to the center of the dipole (in the transverse plane), and $z$ is the fraction of the large photon minus momentum carried by the quark. 
The photon virtuality is $Q^2$, and $\xpom = (M^2+Q^2)/(W^2+Q^2)$ can be interpreted as the fraction of the target longitudinal momentum carried by the pomeron in the infinite momentum frame, with $M$ referring to the vector meson mass and $W$ to the photon-nucleon center-of-mass energy.

The necessary inputs to evaluate the scattering amplitude~\eqref{eq:jpsi_amp} are the photon-vector meson wave function overlap $\Psi_V^*\Psi_\gamma$ and the dipole-target scattering amplitude $N_\Omega$. 
For the non-perturbative vector meson wave function, we apply the Boosted Gaussian model from Ref.~\cite{Kowalski:2006hc}, where the invariant mass distribution for the $q\bar q$ pair is required to be Gaussian around the \jpsi mass. 
In the mixed transverse coordinate-longitudinal momentum fraction space, the wave function overlap reads
\begin{align}
\label{eq:wf_overlap_BG}
    [\Psi_V^* \Psi_\gamma]&(Q^2,\rt,z) = \hat{e}_c e \frac{N_c}{\pi z (1 - z)}\notag \\
    &\times \left\{ m_c^2 K_0(\varepsilon r) \phi_{\rm T}(r, z) \right.
    \notag \\ & ~~~~~~\left. - \left[ z^2 + (1 - z)^2 \right] \varepsilon K_1(\varepsilon r) \partial_r \phi_{\rm T}(r, z) \right\}. 
\end{align}
Here $\hat{e}_c$ is the fractional electric charge of the charm quark, $N_c$ is the number of colors, $\varepsilon^2 = Q^2z(1-z)+m_c^2$, and $m_c$ is the charm quark mass. 
$K_0$ and $K_1$ are Bessel functions of the second kind.
This form is obtained by assuming that the vector meson wave function has the same helicity and polarization structure as the photon wave function (see also related discussions in Ref.~\cite{Lappi:2020ufv,Mantysaari:2020lhf}).
The scalar part $\phi_{\rm T}(\rt,z)$ is described by the Boosted Gaussian model with the parameters fitted to the decay width data~\cite{Kowalski:2006hc}. 
In this work, we use the charm quark mass $m_c=1.4~\mathrm{GeV}$.

The dipole-target scattering amplitude is obtained by combining the McLerran-Venguopalan (MV) model~\cite{McLerran:1993ni} at $\xpom=0.01$ with the JIMWLK-evolution equation~\cite{Mueller:2001uk} that describes the $\xpom$ dependence. 
The dipole-target amplitude is a two-point function of Wilson lines,
\begin{align}
    N_{\Omega}&(\rt,\bt,\xpom)  =\notag\\  & 1 - \frac{1}{\nc} \tr \left[ V\left(\bt + \frac{\rt}{2}\right) V^\dagger\left(\bt - \frac{\rt}{2}\right) \right]. 
\end{align}
The Fourier conjugate to the momentum transfer $\Deltat$ in Eq.~\eqref{eq:jpsi_amp} is the center-of-mass of the dipole.
Note that the Wilson lines $V(\xt)$ depend implicitly on $\xpom$.

In the MV model applied at the initial $\xpom=0.01$, the color charges in the target are assumed to be random local Gaussian variables with zero mean and a two-point correlator
\begin{align}
    g^2 &\langle \rho^a(x^-,\xt) \rho^b(y^-,\yt)\rangle = \notag \\ & \delta^{ab} \delta^{(2)}(\xt-\yt) \delta(x^- - y^-) g^4 \lambda_A(x^-,\xt).
\end{align}
Here the color charge density is $\mu^2(\bt) = \int \dd{x^-} \lambda_A(x^-,\bt)$.
It is related to the local saturation scale $Q_s^2(\bt)$ obtained from the IPsat parametrization~\cite{Kowalski:2003hm,Rezaeian:2012ji}. 
This relation, parametrized as a ratio
\begin{equation}
    \frac{Q_s(\bt)}{g^2\mu(\bt)}\,,
\end{equation}
is treated as a free parameter in the model, controlling the overall proton density. 
Note that a smaller value corresponds to a denser proton. 

The saturation scale $Q_s^2(\bt)$ extracted from the IPsat model depends on the thickness function $T(\bt)$. 
As one of the modifications to the previous iterations of our model, we allow the proton transverse shape to deviate from a Gaussian, and write it as
\begin{equation}
    T_p(\bt) = \Gamma\left(\frac{1}{\omega},\frac{\mathbf{b}_\perp^2}{B_G \omega}\right)/\Gamma\left(\frac{1}{\omega}\right),
    \label{eq:b-profile-proton}
\end{equation}
where $\Gamma(s, x)$ is the upper incomplete gamma function.
For $\omega=1$ we get
\begin{equation}
    T_p(\bt) = \frac{1}{2\pi B_G} e^{-\mathbf{b}_\perp^2/(2B_G)}.
\end{equation}
The more flexible parametrization~\eqref{eq:b-profile-proton}  is motivated by the recent analysis of Ref.~\cite{Lappi:2023frf} showing that the diffractive structure function data from HERA prefers $\omega>1$,~i.e., a density profile that is more steeply falling than a Gaussian. 

Rather than using a smooth nucleon thickness function, we construct it by summing over the density profiles of the nucleon constituents, in this work, $N_q$ hot spots.
It is the hot spot positions $\bti$ whose distribution follows the nucleon density profile~\eqref{eq:b-profile-proton}. 
We sample $N_q$ positions and construct the total proton thickness function by summing the hot spot density profiles whose width is controlled by the parameter $B_q$:
\begin{align}
    T(\bt)&= \sum_{i=1}^{N_q} p_i T_q(\bt-\bti)\,, \quad \text{with} \\
    T_q(\bt) &= \frac{1}{2\pi B_{q} N_q} e^{-\bt^2/(2B_q)}.
\end{align}
This means that the hot spot profile is always assumed to be Gaussian, independent of the proton shape parametrized in Eq.~\eqref{eq:b-profile-proton}.

The overall density ($Q_s^2$) fluctuations are implemented for each hot spot independently. 
In practice, the normalization of the thickness function for the hot spot $i$ is modified by a factor $p_i$ sampled from a log-normal distribution
\begin{equation}
\label{eq:qsfluct}
P\left( \ln p_i \right) = \frac{1}{\sqrt{2\pi}\sigma} \exp \left[- \frac{\ln^2 p_i}{2\sigma^2}\right]\,,
\end{equation}
with $\sigma$ being a parameter controlling the amount of normalization fluctuations.
The sampled values are normalized by the expectation value $e^{\sigma^2/2}$ to obtain $\langle p_i\rangle=1$.
In the case of a nuclear target, we sample nucleon positions from a Woods-Saxon distribution and sum the nucleon thickness functions $T(\bt)$.

Once the color charge distribution is obtained for the proton or a nucleus, Wilson lines are computed by solving the Yang-Mills equations. 
In terms of the color charge distributions $\rho$, one obtains
\begin{equation}
    V(\xt) = P_- \exp \left\{ -ig \int \dd{x^-} \frac{\rho^a(x^-,\xt) t^a}{\nabla^2_{\xt} -  m^2} \right\}.
\end{equation}
Here, the infrared regulator $m$ controlling the long-distance Coulomb tails is a free parameter of the model.

We also consider a modification to the MV model that suppresses the high-frequency modes (scattering amplitude for small dipoles) following Ref.~\cite{Mantysaari:2018zdd}. 
This is motivated by the fact that leading-order dipole model fits to structure function data prefer dipole-proton scattering amplitudes that decrease faster than the MV model prediction $\sim r^2$ as $r$ goes to $0$~\cite{Albacete:2010sy,Lappi:2013zma}. 
In practice, this is achieved by introducing a ultraviolet (UV) damping factor $v_\mathrm{UV}$ as
\begin{equation}
\label{eq:uvdamp}
    A^+(x^-, \mathbf{k}) = -\frac{\rho(x^-, \kt)}{\kt^2 + m^2} e^{-|\kt| v_\mathrm{UV}}.
\end{equation}
Here $\rho(x^-,\kt)=\rho^a(x^-,\kt)t^a$ is the color charge distribution in transverse Fourier space.
The MV model is recovered in the limit $v_\mathrm{UV}\to 0$.

The Wilson lines are evolved to smaller $\xpom$ by solving the JIMWLK evolution equation event-by-event~\cite{Mueller:2001uk}. 
The numerical implementation, publicly available in the codes~\cite{jimwlk_code,ipglasma_jimwlk_code}, follows Ref.~\cite{Lappi:2012vw}.
For a detailed discussion of the numerical solution to JIMWLK, see Ref.~\cite{Cali:2021tsh}.
We use the daughter-dipole running coupling prescription, and the strong coupling constant in the coordinate space is parametrized as
\begin{equation}
    \as(r) = \frac{12\pi}{(11\nc - 2\nf) \ln \left[ \left(\frac{\mu_0^2}{\lqcd^2}\right)^{1/\zeta} + \left(\frac{4}{r^2\lqcd^2}\right)^{1/\zeta} \right]^\zeta}.
\end{equation}
Here, $r$ are the dipole sizes of the daughter-dipoles appearing in the JIMWLK equation. 
The parameter $\lqcd$ in the coordinate space is considered as a free parameter controlling the scale as a function of transverse separation, and the parameters regulating the Landau pole are $\mu_0=0.28~\gev$ and $\zeta=0.2$. 
Furthermore, we set $\nf=3$. 

To regulate Coulomb tails and avoid an unphysically fast growth of the nucleon size, the JIMWLK kernel describing gluon emission to distance $\xt$ has to be regulated. 
We follow Ref.~\cite{Schlichting:2014ipa} and write the JIMWLK kernel $K^i$ as
\begin{equation}
\label{eq:jimwlk_kernel}
    K^i({\xt}) = \mjimwlk |\xt| K_1(\mjimwlk |\xt|) \frac{x^i}{{\mathbf x}_\perp^2}.
\end{equation}
In the limit $\mjimwlk \to 0$, this reduces back to the perturbative expression $K^i(\xt)=x^i/{\mathbf x}_\perp^2$.

\section{Bayesian inference setup}
\label{sec:Bayesian_inference}
We employ Bayesian inference~\cite{10.1093/oso/9780198568315.001.0001} as a systematic framework to constrain the probability distributions of the model parameters $\btheta$ by comparing model predictions $\by(\btheta)$ with experimental measurements $\by_{\rm exp}$. 
This approach also enables the propagation of model uncertainties to the final predictions for the cross sections.

Bayes' theorem is given by:
\begin{equation}
    \mathcal{P}(\btheta|\mathbf{y}_{\rm exp}) = \frac{\mathcal{P}(\mathbf{y}_{\rm exp}|\btheta)\mathcal{P}(\btheta)}{\mathcal{P}(\mathbf{y}_{\rm exp})},
\end{equation}
where $\mathcal{P}(\btheta|\mathbf{y}{\rm })$ is the posterior distribution of the parameters, $\mathcal{P}(\mathbf{y}_{\rm exp}|\btheta)$ is the likelihood, $\mathcal{P}(\btheta)$ is the prior, and $\mathcal{P}(\mathbf{y}_{\rm exp})$ is the evidence.
We choose uniform distributions for the model prior $\mathcal{P}(\btheta)$ in every dimension and assume a multi-variant Gaussian distribution for the likelihood function:
\begin{align}
    &\mathcal{P}(\mathbf{y}_{\rm exp}|\btheta) = \frac{1}{\sqrt{2\pi |\mathrm{det}(\Sigma)|}} \nonumber \\
    & ~~~~ \times\exp\left[-\frac{1}{2} (y(\btheta) - y_\mathrm{exp})^\mathsf{T} \Sigma^{-1} (y(\btheta) - y_\mathrm{exp})\right].
\end{align}
Here, the covariance matrix includes uncertainties from both experimental measurements and the model, $\Sigma = \Sigma_\mathrm{model} + \Sigma_\mathrm{exp}$.
The model uncertainty is estimated from the accuracy of the GP emulator.
We assume uncorrelated experimental errors for our inference analysis since most of the experimental data does not provide the covariances explicitly.\footnote{To be consistent with the experimental covariance matrix, we will also use the diagonal components of the emulator's covariance matrix only. The possibility of including the only available covariance matrix for the integrated $\gamma+\mathrm{Pb}$ data will be discussed later.}
The posterior distribution and the Bayesian evidence are obtained using the \texttt{pocoMC} sampler~\cite{Karamanis:2022alw,Karamanis:2022ksp}.

Due to the high computational cost of simulating $\gamma+\mathrm{Pb}$ and $\gamma+p$ cross sections, and the requirement to evaluate the likelihood at many points in parameter space, we use Gaussian Process (GP) emulators~\cite{Rasmussen2006Gaussian} to approximate the forward model during sampling.
Following similar approaches as in Refs.~\cite{Roch:2024xhh,Jahan:2024wpj,Jahan:2025cbp}, we utilize a Bayesian analysis package~\cite{hendrik_roch_2025_15879411} that interfaces with various GP emulators and the \texttt{pocoMC} sampler.
For this study, we employ two types of emulators: the PCGP emulator from the \texttt{surmise} package developed by the BAND collaboration~\cite{surmise2023}, and the standard GP implementation from the Scikit-learn Python package~\cite{scikit-learn}. 
The latter has been widely used in previous Bayesian analyses in the field of high-energy nuclear physics~\cite{Bernhard:2019bmu,JETSCAPE:2020mzn,Nijs:2020ors,Parkkila:2021tqq,Mantysaari:2022ffw,Heffernan:2023gye,Soeder:2023vdn,Shen:2023awv,Shen:2023pgb}.

Table~\ref{tab:parameters} summarizes the model parameters and their associated prior ranges together with the estimated MAP parameters obtained from the 25 posterior sample runs with the smallest $\chi^2/\mathrm{dof}$ for the model with fixed and variable $K$ separately.\footnote{To improve the determination of the MAP parameters when including the $\Kwf$ factor, we varied $\Kwf$ for each of the 25 posterior samples while keeping all other parameters fixed, and selected the value that yields the best fit to the experimental data.}
\begin{table*}[tb!]
    \caption{Summary of model parameters and their prior ranges. The first block corresponds to the standard setup from Refs.~\cite{Mantysaari:2022sux,Mantysaari:2023xcu}, the second to JIMWLK evolution parameters, and the third to model extensions investigated in Sec.~\ref{sec:extensions}. The last two columns show the estimated model MAP parameter sets from the 25 posterior sample runs with the smallest $\chi^2/\mathrm{dof}$ compared to the experimental data.
    }
    \label{tab:parameters}
    \begin{tabular}{l|l|l|l|l|l}
    \hline\hline
    Block & Parameter & Description & Prior range & MAP ($K\equiv 1$) & MAP (variable $K$) \\
    \hline
    1 & $m\;[\mathrm{GeV}]$ & Infrared regulator &  $[0.02,1.2]$ & 0.31 & 0.51 \\
    & $B_G\;[\mathrm{GeV}^{-2}]$ & Proton size & $[1,10]$ & 2.99 & 3.83 \\
    & $B_{q}\;[\mathrm{GeV}^{-2}]$ & Hot spot size & $[0.05,3]$ & 0.07 & 0.30 \\
    & $\sigma$ & Magnitude of $Q_s$ fluctuations & $[0,1.5]$ & 0.99 & 0.88 \\
    & $Q_s/(g^2\mu)$ & Ratio of color charge density to saturation scale & $[0.05,1.5]$ & 0.63 & 0.37 \\
    \hline
    2 & $m_{\mathrm{JIMWLK}}\;[\mathrm{GeV}]$ & Infrared regulator & $[0.02,1.2]$ & 0.12 & 0.16 \\
    & $\Lambda_{\mathrm{QCD}}\;[\mathrm{GeV}]$ & Spatial $\Lambda_{\rm QCD}$ & $[0.0001,0.28]$ & 0.0097 & 0.0348 \\
    \hline
    3 & $N_q$ & Number of hot spots & $[0,10]$ & 3 (fixed) & 3 (fixed)\\
    & $\omega$ & Modification of the proton shape & $[0,10]$ & 1 (fixed) & 1 (fixed) \\
    & $v_\mathrm{UV}\;[\mathrm{GeV}^{-1}]$ & Damping of high-frequency modes & $[0.0,1.0]$ & 0 (fixed) & 0 (fixed) \\ 
    & $K$ & Cross section scaling factor & $[0.01,4]$ & 1 (fixed) & 0.308 \\
    \hline
    & & & & $\chi^2/\mathrm{dof}=5.94$ & $\chi^2/\mathrm{dof}=1.82$ \\
    \hline\hline
    \end{tabular}
\end{table*}
The first two blocks correspond to the ``standard'' parameter setup used also in,~e.g., Refs.~\cite{Mantysaari:2022sux,Mantysaari:2023xcu}, and the third block contains parameters introduced to explore model extensions in this work. 
In the first block, we adopt a baseline setup similar to Ref.~\cite{Mantysaari:2022ffw}.\footnote{We omit the minimum 3D distance between hot spots, as it could not be constrained with the observables considered in Ref.~\cite{Mantysaari:2022sux} as shown in Ref.~\cite{Mantysaari:2022ffw}.} 
The second block contains the JIMWLK parameters.
The parameters in the third block vary the average proton shape, sub-nucleonic structure, and fluctuations. 
The $K$ factor is a multiplicative rescaling of all cross sections, as shown in Eqs.~\eqref{eq:coh} and~\eqref{eq:incoh}.

To train the surrogate GP emulators (for $\gamma+\mathrm{Pb}$ and $\gamma+p$ integrated and differential cross sections), we adopt an iterative strategy for generating training data. 
We begin with a maximum-projection Latin Hypercube Design (LHD)~\cite{joseph2015maximum} and evaluate the full model at approximately 850 design points for each collision system.\footnote{The number of design points varies slightly between datasets for different systems and observables.}
We supplement this with an additional LHD of approximately 170 points for $v_\mathrm{UV}\equiv 0$ to ensure enough training coverage for our standard model setup, which is defined at this fixed value of $v_\mathrm{UV}$.

Because the JIMWLK evolution depends logarithmically on the parameter $\Lambda_{\mathrm{QCD}}$, we assign the LHD training points to $\log(\Lambda_{\mathrm{QCD}})$ and also add approximately 450 design points sampled for $\log(\mu_0 - \Lambda_{\mathrm{QCD}})$, which populate the parameter phase-space more densly when $\Lambda_{\mathrm{QCD}}$ is near the Landau pole. 
These additional training points are effective to reduce emulator uncertainties for the integrated cross sections when $\Lambda_{\mathrm{QCD}} \sim \mu_0$.

Next, we perform an active-learning approach to reduce the emulator uncertainties over the entire prior parameter space~\cite{JETSCAPE:2024cqe}. 
We employ a greedy-algorithm-based approach~\cite{VINCE2002247} to identify the phase space where the trained emulators have relatively large uncertainties and generate training points in these regions. 
We added 200 training points in the first iteration and another 240 points in the second iteration.

Finally, to further enhance the emulator accuracy in the physically relevant region of parameter space, we perform an intermediate Bayesian inference and sample 50 additional training points from the resulting posterior distribution. This final step significantly improves emulator accuracy in the high-likelihood region.

As mentioned above, in this work we employ the PCGP emulator from the \texttt{surmise} Python package~\cite{surmise2023} as a fast surrogate model for Bayesian inference.
It performs Principal Component Analysis (PCA) on the training data to reduce the number of required GPs. 
The PCGP emulator demonstrates the highest precision and most reliable uncertainty quantification among all the types of GPs available to us.
We adopted it for modeling the $\gamma+\mathrm{Pb}$ datasets and the integrated $\gamma+p$ cross section.

For the differential $\gamma+p$ cross sections, however, the situation is more nuanced.
The training data contains highly non-uniform statistical uncertainties for the $\gamma+p$ cross sections at different $|t|$ values. 
The relative statistical errors are large for coherent differential cross sections at high $|t|$ and incoherent cross sections at small $|t|$. 
In this case, the use of principal component analysis combines observables with varying uncertainty scales, resulting in the GP overestimating uncertainties for observables with small statistical uncertainties in the training data. 

Since PCA cannot be disabled in the PCGP implementation, we instead employ the Scikit-learn GP emulator for the differential $\gamma+p$ cross sections and train individual data points independently.
This approach preserves the native structure of the data and improves the emulator's accuracy in this challenging regime.

To quantitatively assess the emulator's performance across multiple closure tests, we adopt an information-theoretic metric introduced in Ref.~\cite{JETSCAPE:2023ikg} and further refined in Ref.~\cite{Roch:2024xhh}. 
This metric is defined as
\begin{equation}
\Delta \equiv \frac{1}{N_{\rm param.}}\int \left|\frac{\btheta-\btheta_{\rm truth}}{\btheta_{\rm max}-\btheta_{\rm min}}\right|^2 \mathcal{P}(\btheta|\mathbf{y}(\btheta_{\rm truth}))\;\mathrm{d}\btheta,
\label{eq:delta_d}
\end{equation}
which measures the average squared deviation of the posterior $\mathcal{P}(\btheta|\mathbf{y}(\btheta_{\rm truth}))$ conditioned by the pseudodata generated at the true parameter value $\btheta_{\rm truth}$, normalized by the prior range $|\btheta_{\rm max}-\btheta_{\rm min}|$ and the number of parameters $N_{\rm param.}$.
Smaller values of $\Delta$ indicate posteriors that are more sharply peaked around the true parameter values, implying better emulator precision and accuracy.

In our analysis, we compute $\Delta$ for a total of $N_{\rm samples} = 50$ posterior distributions in the full 11-dimensional parameter space, including 25 parameter sets with $K \equiv 1$ and 25 where $K$ is varied. 
We report the average value
\begin{equation}
\langle\Delta\rangle = \frac{1}{N_{\rm samples}} \sum_{i=1}^{N_{\rm samples}} \Delta_i,
\end{equation}
which summarizes the overall emulator performance across all closure tests $i$. 
We obtain $\langle\Delta\rangle = 0.040 \pm 0.002$.
In comparison, a random guess from a uniform distribution would yield $\langle\Delta\rangle = 1/6$. 
If the posterior distribution is a multivariate Gaussian centered at the truth, the value $\langle\Delta\rangle = 0.040 \pm 0.002$ corresponds to a normalized standard deviation $\sigma \sim 0.2$~\cite{JETSCAPE:2023ikg}. 
Hence, our closure test result indicates that the trained GP emulators are precise enough to constrain the model parameters within 20\% relative uncertainties.\footnote{Decreasing the emulator uncertainty further would require much more computation time, especially for the $|t|$-differential $\gamma+$p cross section, which requires averaging over many target configurations $\Omega$.}

\section{Results}
\label{sec:results}
In this section, we present results from the Bayesian analysis, including the extracted posterior distributions of model parameters, and model comparisons to the experimental data based on sampling the parameter posteriors.
This procedure allows us to assign a systematic theory uncertainty to the calculated observables.
In previous analyses~\cite{Mantysaari:2023xcu,Mantysaari:2022sux} the non-perturbative parameters have been fitted to the 
$\gamma+p\to\jpsim+p$ data and a successful description of the experimental measurements across a wide range in center-of-mass energy $W$ has been obtained. 
On the other hand, such calculations typically overestimate the $\gamma+\mathrm{Pb}\to\jpsim+\mathrm{Pb}$ cross sections, especially at high $W$.
Here, we aim to determine if including the available nuclear data in the analysis and performing a Bayesian analysis covering the full parameter space systematically can resolve this tension. 
Furthermore, we will explore whether the more flexible initial state description introduced in this work can help alleviate some of the tension.

In the analysis, we include the following datasets:
\begin{itemize}
    \item $\gamma+p\to\jpsim+p$ coherent cross section integrated over $t$ in the range $50.4<W<1579\;\gev$, using datasets from H1~\cite{H1:2005dtp,H1:2013okq}, ZEUS~\cite{ZEUS:2002wfj}, ALICE~\cite{ALICE:2014eof,ALICE:2018oyo}, and LHCb~\cite{LHCb:2018rcm}.
    \item $\gamma+p\to\jpsim+p$ coherent and incoherent cross sections differential in $t$ at $W=75\;\gev$, using the dataset from H1~\cite{H1:2013okq}.
    \item $\gamma+\mathrm{Pb}\to\jpsim+\mathrm{Pb}$ coherent cross section integrated over $t$ in the range $31.5<W<813\;\gev$, using datasets from ALICE~\cite{ALICE:2023jgu} and CMS~\cite{CMS:2023snh}.
    \item $\gamma+\mathrm{Pb}\to\jpsim+\mathrm{Pb}^{(*)}$ coherent and incoherent  cross sections differential in $t$ at $W=125\;\gev$, using the dataset from ALICE from Refs.~\cite{ALICE:2021tyx} (coherent) and~\cite{ALICE:2023gcs} (incoherent).
\end{itemize}

The cross sections for the $\gamma+\mathrm{Pb}\to\jpsim+\mathrm{Pb}$ scattering have been extracted from the ultra peripheral $\mathrm{Pb}+\mathrm{Pb}\to\jpsim+\mathrm{Pb}+\mathrm{Pb}$ data by the experimental collaborations. 
This extraction relies on the method proposed in Ref.~\cite{Guzey:2013jaa}, and naturally results in a large anticorrelation between the low-$W$ and high-$W$ results. 
Including the full experimental covariance matrix for the $\gamma+\mathrm{Pb}$ integrated cross sections was found to have little effect on the outcome of the fit, because this cross section has a small relative weight in the likelihood compared to the rest of the experimental measurements, which do not have a cross correlation with each other. 

We note that also preliminary ATLAS data exists~\cite{ATLAS:2025uxr} for the \jpsi photoproduction in UPCs. 
However, this data is still preliminary, and ATLAS has not yet extracted the $\gamma+\mathrm{Pb}$ cross section.\footnote{At midrapidity, the ATLAS data is not compatible with the ALICE data. A robust Bayesian inference study requires this discrepancy to be resolved.}
Thus, we do not include that dataset in this analysis.

\subsection{Standard setup}
\label{sec:standardsetup}
\begin{figure*}[t]
      \includegraphics[width=\textwidth]{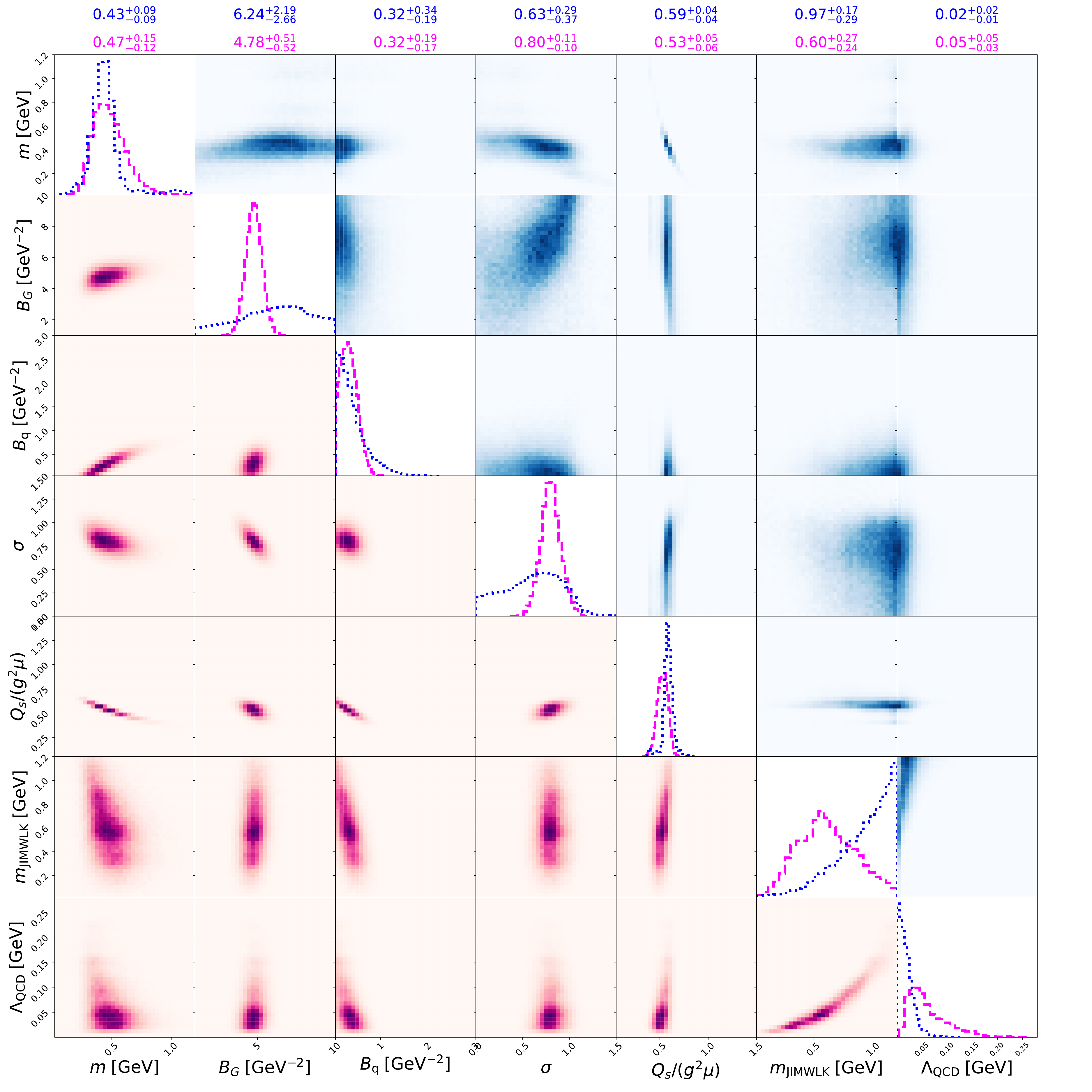}
    \caption{Posterior distributions obtained by fitting the $\gamma+p$ (dashed, lower corner) and $\gamma+\mathrm{Pb}$ (dotted, upper corner) data separately. The numbers presented at the top of the figure are the median values along with their corresponding 90\% credible intervals.} 
    \label{fig:posterior_defaultsetup}
\end{figure*}

In the first analysis, we use the same setup as in Refs.~\cite{Mantysaari:2023xcu,Mantysaari:2022sux}. 
This corresponds to not introducing any additional normalization factor (fixing $\Kwf\equiv1$), using the MV model without filtering high-frequency modes ($v_\mathrm{UV}\equiv 0$), using $N_q\equiv 3$ hot spots, and employing a Gaussian proton shape ($\omega\equiv 1$).
In order to quantify the tension between the $\gamma+p$ and $\gamma+\mathrm{Pb}$ datasets observed in Refs.~\cite{Mantysaari:2023xcu,Mantysaari:2022sux}, we first perform two separate Bayesian analyses by including only either the $\gamma+p$ or $\gamma+\mathrm{Pb}$ data. 
The obtained posterior distributions are shown in Fig.~\ref{fig:posterior_defaultsetup}.
The global analysis, including both datasets simultaneously, is presented in Sec.~\ref{sec:kfact_analysis}.

In the $\gamma+p$ case (lower part of Fig.~\ref{fig:posterior_defaultsetup} and dashed magenta curves on the diagonal), the obtained posterior distribution is compatible with the previous fit reported in Ref.~\cite{Mantysaari:2022sux}.\footnote{In Ref.~\cite{Mantysaari:2022sux}, a slightly different parametrization for the \jpsi wave function was used with a different charm quark mass, which mostly affects the overall normalization controlled by $Q_s/(g^2\mu)$.}
The available $\gamma+p$ data constrains all model parameters tightly, except the IR regulator $m_\mathrm{JIMWLK}$ in the JIMWLK kernel~\eqref{eq:jimwlk_kernel} and the scale $\Lambda_{\mathrm{QCD}}$ in the running coupling, for which the obtained distributions are relatively broad and a clear correlation between the two parameters is found, similarly to Ref.~\cite{Mantysaari:2018zdd}. 

Strong constraints for the parameters are obtained because different parameters have different sensitivities to the various features of the exclusive meson production data~\cite{Mantysaari:2022ffw}. 
In particular, the behavior of the coherent cross section at low $|t|$ is sensitive to the target geometry at large impact parameters, where Coulomb tails regulated by $m$ dominate~\cite{Mantysaari:2016jaz}. 
The slope of the coherent cross section probes the proton size controlled most directly by $B_{G}$, and the slope of the incoherent cross section probes the hot spot size $B_q$~\cite{Lappi:2010dd,Demirci:2022wuy}. 
The incoherent cross section at low $|t|$, on the other hand, is mostly sensitive to the $Q_s$ fluctuations parametrized by $\sigma$~\cite{Mantysaari:2016jaz}. 

The $\gamma+\mathrm{Pb}$ data by itself poses weaker constraints on most parameters compared to the $\gamma+p$ data (upper part of Fig.~\ref{fig:posterior_defaultsetup} and blue dotted lines on the diagonal). 
Still, relatively strong constraints are obtained for the IR regulator $m$ in the MV model, the hot spot size $B_q$, and the overall saturation scale $Q_s/(g^2\mu)$. 
We note that while the available incoherent $\gamma+\mathrm{Pb}\to\jpsim+\mathrm{Pb}^*$ cross section data at relatively high $|t|$ provides strong constraints on the hot spot size (compatible with the $\gamma+p$ analysis), only weak constraints on the nucleon size and the magnitude of $Q_s$ fluctuations ($\sigma$) are obtained.
The parameters related to the JIMWLK evolution ($m_\mathrm{JIMWLK}$ and $\lqcd$) are not well constrained and are again found to be strongly correlated. 

Some differences between nucleon structure parameters are found when fitting to $\gamma+\mathrm{Pb}$ data as compared to $\gamma+p$ data. While a somewhat larger nucleon size, parametrized by $B_G$ is preferred in the $\gamma+\mathrm{Pb}$ case, the median hot spot size $B_q$ is very similar. We note that when performing another analysis (not shown), in which we allow $N_q$ to vary, the posterior is peaked around $N_q\equiv 3$ when fitting to $\gamma+p$ and to $\gamma+\mathrm{Pb}$ data. A more detailed study of whether the data supports nuclear effects on the nucleon structure beyond those obtained from JIMWLK evolution is left for future work.

\begin{figure*}[tb]
  \centering
  \subfloat[$\gamma+p\to\jpsim+p$ cross section as a function of center-of-mass energy compared to ALICE~\cite{ALICE:2014eof,ALICE:2018oyo}, H1~\cite{H1:2005dtp,H1:2013okq}, ZEUS~\cite{ZEUS:2002wfj}, and LHCb~\cite{LHCb:2018rcm,LHCb:2024pcz} data. Datapoints with open markers are not included in the fit.]{%
    \includegraphics[width=0.48\textwidth]{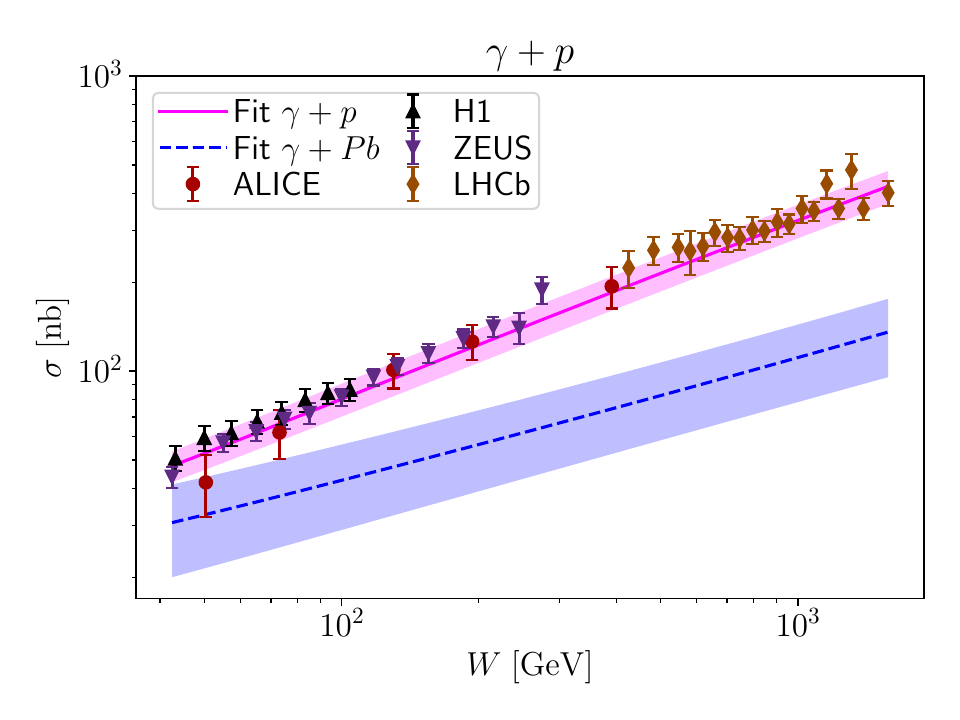}%
    \label{fig:gammap_separatefits}%
  }
  \hfill
  \subfloat[$\gamma+\mathrm{Pb}\to\jpsim+\mathrm{Pb}$ cross section as a function of center-of-mass energy compared to ALICE~\cite{ALICE:2023jgu} and CMS~\cite{CMS:2023snh} data.]{%
    \includegraphics[width=0.48\textwidth]{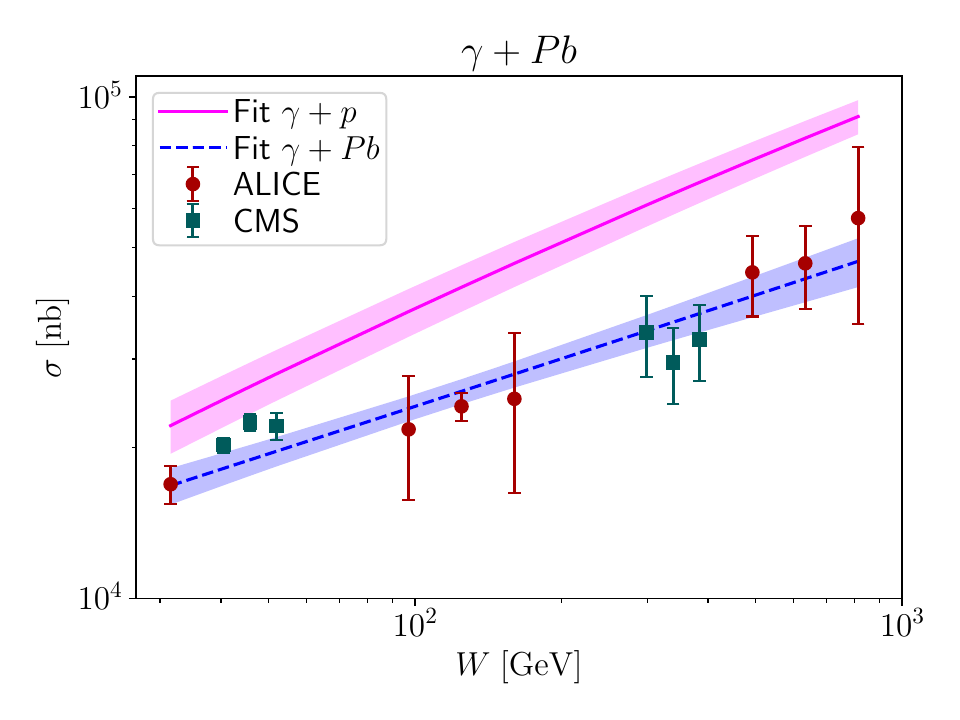}%
    \label{fig:gammaPb_separatefits}%
  }
  \hfill
  \subfloat[Coherent (smaller $|t|$) and incoherent (larger $|t|$) \jpsi spectra in $\gamma+p$ compared to H1 data~\cite{H1:2013okq}.]{%
    \includegraphics[width=0.48\textwidth]{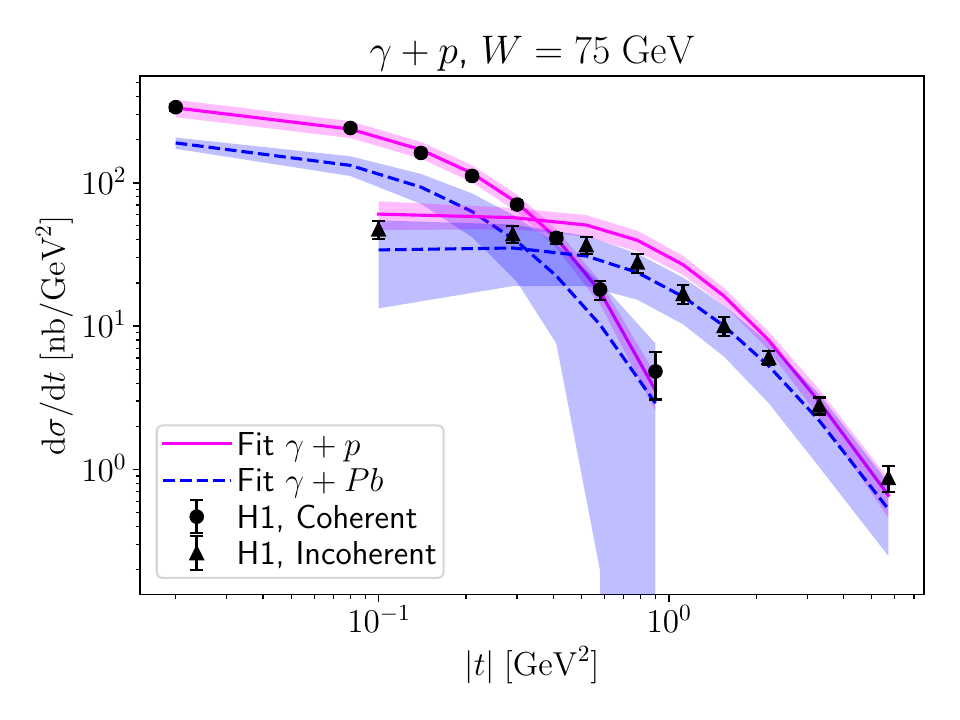}%
    \label{fig:gammap_separatefits_t}%
  }
  \hfill
  \subfloat[Coherent (smaller $|t|$) and incoherent (larger $|t|$) \jpsi spectra in $\gamma+\mathrm{Pb}$ compared to ALICE data~\cite{ALICE:2021tyx,ALICE:2023gcs}.]{%
    \includegraphics[width=0.48\textwidth]{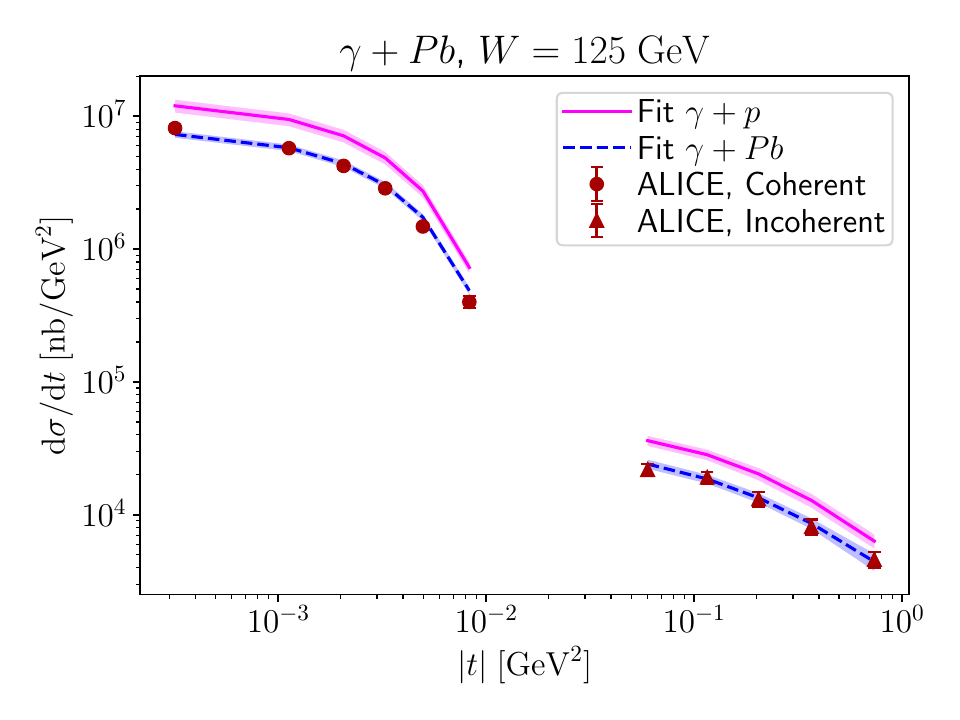}%
    \label{fig:gammaPb_separatefits_t}%
  }
  \caption{Integrated $W$ dependent and $|t|$-differential cross sections from two separate fits containing only $\gamma+p$ (full), or only $\gamma+\mathrm{Pb}$ (dashed) data using the standard parameter setup.
  The uncertainty bands indicate the 68\% credible intervals from 25 posterior sample runs of the model.}
  \label{fig:combined_separatefits}
\end{figure*}

Comparisons of our results as a function of center-of-mass energy $W$, obtained by averaging full model simulations over 25 posterior samples, to the $\gamma+p\to\jpsim+p$ and $\gamma+\mathrm{Pb}\to\jpsim+\mathrm{Pb}$ data are shown in Figs.~\ref{fig:gammap_separatefits} and~\ref{fig:gammaPb_separatefits}. 
Similarly to Ref.~\cite{Mantysaari:2022sux}, one can successfully fit the $\gamma+p\to\jpsim+p$ data, but when the fit is applied to predict the cross section with lead targets, the available data is overestimated, especially at $W\gtrsim 100$ GeV.
On the other hand, the $\gamma+\mathrm{Pb}$ fit can successfully describe the available nuclear data, but when applied to $\gamma+p\to\jpsim+p$, the cross section is significantly underestimated at large $W$. 
This tension arises because the two datasets seem to favor different rates of energy evolution.

We observe similar trends for the $|t|$-differential coherent and incoherent cross sections shown in Figs.~\ref{fig:gammap_separatefits_t} and~\ref{fig:gammaPb_separatefits_t}.
The fit to $\gamma+p\to \mathrm{J}/\psi+p$ data reproduces the coherent $\gamma+p$  cross section well. 
However, it slightly overestimates the incoherent cross section at small $|t|$, likely due to inaccuracies in the fit from larger emulator uncertainties in this region.

When applying the same fit to the nuclear data, we find a notable overestimation of the $|t|$-differential cross sections.
Conversely, the fit performed directly on the nuclear data describes the measured $|t|$-differential $\gamma+\mathrm{Pb}$ spectra well, but it underestimates the $\gamma+p\to \mathrm{J}/\psi+p$ cross section. This is consistent with the behavior seen in the integrated cross sections discussed earlier.
Interestingly, the incoherent $|t|$-differential $\gamma+p\to \mathrm{J}/\psi+p$ cross section is captured reasonably well by the fit to the nuclear data, which might be a coincidence due to increasing emulator uncertainties in this region. 

\subsection{Model extensions}
\label{sec:extensions}
We defined the simplest model considered in the previous subsection, which includes the first seven parameters listed in Table~\ref{tab:parameters}, as the \textit{standard} setup.
To evaluate whether additional non-perturbative parameters are supported by the data, we extend this standard model with one or more such parameters and compare the resulting models using the Bayes factor. 
This quantity compares the posterior probabilities of two models $A$ and $B$ given the experimental observations $\mathbf{y}_{\rm exp}$ (in this subsection we include both $\gamma+p$ and $\gamma+{\rm Pb}$ data):
\begin{align}
    \mathcal{B}_{A/B} \equiv \frac{\mathcal{P}(A \lvert \mathbf{y}_{\rm exp})}{\mathcal{P}(B \lvert \mathbf{y}_{\rm exp})} = \frac{\mathcal{P}(\mathbf{y}_{\rm exp}\lvert A)\mathcal{P}(A)}{\mathcal{P}(\mathbf{y}_{\rm exp}\lvert B)\mathcal{P}(B)}.
    \label{eq:B_factor}
\end{align}
Assuming there is no bias on prior beliefs about the validity of these two models,~i.e., $\mathcal{P}(A) = \mathcal{P}(B)$, Eq.~\eqref{eq:B_factor} simplifies to
\begin{equation}
    \mathcal{B}_{A/B} = \frac{\mathcal{P}(\mathbf{y}_{\rm exp}\lvert A)}{\mathcal{P}(\mathbf{y}_{\rm exp}\lvert B)},
    \label{eq:B_evidence}
\end{equation}
where the Bayesian evidence for model $A$ is computed via marginalization over its parameter space:
\begin{equation}
    \mathcal{P}(\mathbf{y}_{\rm exp}\lvert A) = \int\mathrm{d}\btheta_{A}\;\mathcal{P}(\mathbf{y}_{\rm exp}\lvert {\btheta}_{A})\mathcal{P}({\btheta}_{A}).
    \label{eq:B_marginalized_L}
\end{equation}
We evaluate the Bayesian evidence using the \texttt{pocoMC} package~\cite{Karamanis:2022alw,Karamanis:2022ksp}. 
For interpreting the Bayes factor, we follow the Jeffreys' scale~\cite{Gordon:2007xm,Trotta:2008qt}, which classifies the strength of evidence based on $\ln\mathcal{B}_{A/B}$. 
In particular, values $\ln\mathcal{B}_{A/B} < -5$ are considered ``strong evidence'' in favor of model $B$.

Table~\ref{tab:Bayes_factors} summarizes the results for various model extensions relative to the standard setup. 
Adding $v_{\rm UV}$ leads to a small preference for the standard setup. 
In contrast, extending the base model with the proton shape parameter $\omega$ and the number of hotspots $N_q$ yields weak evidence in favor of the extended model. 
Including $v_{\rm UV}$ in addition to these parameters slightly reduces the preference, consistent with the first comparison.
Allowing only the number of hotspots $N_q$ to vary is slightly more favored by the data than simultaneously varying both $N_q$ and the proton shape parameter $\omega$.

Introducing the $K$ factor yields strong evidence in favor of the extended model. The table then compares this preferred model to extensions to it, taking the model extended by $K$ to be model A. All values on the Jeffrey's scale are positive for those comparisons, meaning that adding further parameters ($v_{\rm UV}$, $\omega$, $N_q$) reduces the overall model evidence. We conclude that the experimental data prefer the base model extended by the $K$ factor only, while the parameters $v_{\rm UV}$, $\omega$, and $N_q$ can be fixed to their default values of 0, 1, and 3, respectively.

\begin{table}[tb]
    \caption{Bayes factors comparing the standard setup (with seven parameters), or the standard setup extended by the normalization factor $K$, to extended setups that include additional parameters. Negative values of $\ln(\mathcal{B}_{A/B})$ indicate a preference for model $B$ over model $A$.}
    \label{tab:Bayes_factors}
    \centering
    \begin{tabular}{c|c|c}
        \hline\hline
        Model $A$ & Model $B$ & $\ln(\mathcal{B}_{A/B})$ \\
        \hline
        standard & $v_{\rm UV}$ & $0.63\pm 0.02$ \\
        standard & $\omega, N_q$ & $-2.18\pm 0.02$ \\
        standard & $N_q$ & $-3.12\pm 0.02$ \\
        standard & $\omega, v_{\rm UV}, N_q$ & $-1.03\pm 0.02$ \\
        standard & $K$ & $-18.56\pm 0.02$ \\
        $K$ & $K, v_{\rm UV}$ & $3.39\pm 0.02$ \\
        $K$ & $K, N_q$ & $1.02\pm 0.01$ \\
        $K$ & $K, \omega, N_q$ & $1.48\pm 0.02$ \\
        $K$ & $K, \omega, v_{\rm UV}, N_q$ & $4.60\pm 0.02$ \\
        \hline\hline
    \end{tabular}
\end{table}

\subsection{Global analysis with and without a $K$ factor}
\label{sec:kfact_analysis}
Next, we proceed to perform a global analysis using the extended model with the normalization factor $\Kwf$ (that multiplies all cross sections) considered as a free parameter.
As discussed in Sec.~\ref{sec:setup}, this factor can be interpreted, e.g., to parametrize the wave function uncertainty or to capture missing higher-order effects. 
This factor can also be expected to help resolve the tension between the $\gamma+p$ and $\gamma+\mathrm{Pb}$ datasets: if $\Kwf<1$, its effect has to be compensated by a larger color charge density, which results in enhanced non-linear effects and stronger nuclear suppression.\footnote{How the nuclear modification factor in exclusive \jpsi production depends on the wave function model has been quantified in Ref.~\cite{Lappi:2020ufv}.} 
We will also compare to the case of a global analysis with $K\equiv 1$.

\begin{figure*}[t]
      \includegraphics[width=\textwidth]{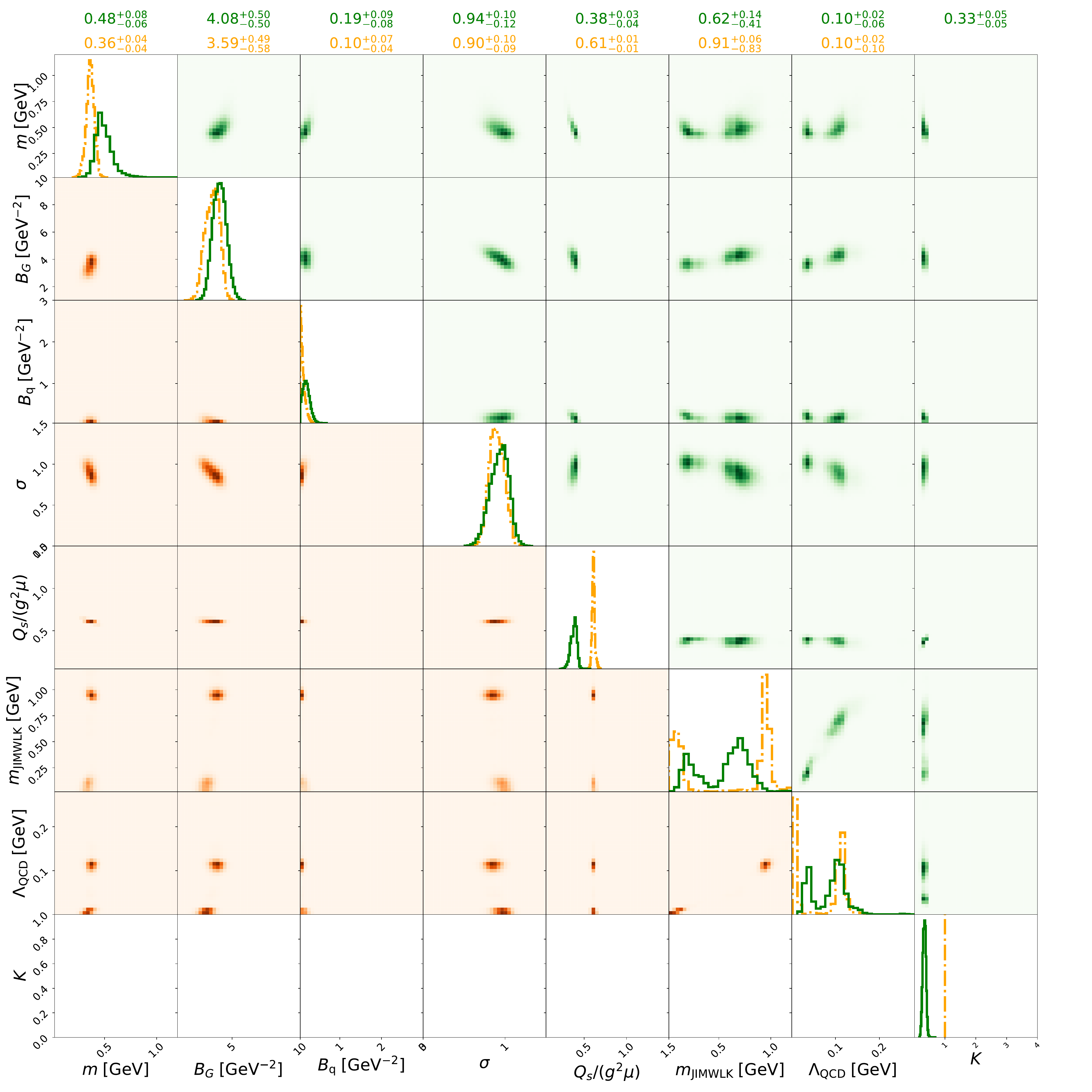}
    \caption{Posterior distribution obtained in the global analysis result with (full, upper corner) and without (dash-dotted, lower corner) the model extension by the normalization factor $\Kwf$. The numbers presented at the top of the figure are the median values along with their corresponding 90\% credible intervals.}
    \label{fig:posteriorwithK}
\end{figure*}

The obtained posterior distributions for both cases are shown in Fig.~\ref{fig:posteriorwithK}. 
The lower half corresponds to the setup with $\Kwf\equiv 1$,~i.e., it is equivalent to the setup discussed in Sec.~\ref{sec:standardsetup}, but now both the $\gamma+p$ and $\gamma+\mathrm{Pb}$ datasets are included in the analysis simultaneously. 
The determined model parameters related to the initial state are approximately compatible with the ones obtained from separate $\gamma+p$ or $\gamma+\mathrm{Pb}$ analyses, but the parameters related to the JIMWLK evolution are still not well constrained. 
Instead, a double-peak structure appears, which we interpret as a result of the tension in the evolution speeds between the two datasets.\footnote{We verified that the observed double-peak structure is not a numerical artifact by employing an alternative MCMC sampler, PTLMC from the \texttt{surmise} package, which yielded consistent posterior distributions. The emulators were further validated by varying the parameters $m_{\rm JIMWLK}$ and $\lqcd$ between the peaks, confirming a local minimum of the likelihood in this region. Full model evaluations showed that the $\chi^2/\mathrm{dof}$ is lower at the peak positions than between them.}

When the normalization factor $\Kwf$ is included, we find a preference for a small value $\Kwf \sim 0.3$. 
We note that the actual numerical value obtained for this parameter would also depend on the chosen charm quark mass (here $m_c=1.4~\mathrm{GeV}$) and on the model chosen for the non-perturbative meson wave function. This reflects the fact that $K$ can be used to absorb some model uncertainties.
As anticipated, the smaller normalization due to $\Kwf<1$ is compensated by a smaller value obtained for the $Q_s/(g^2\mu)$ ratio, which implies a larger color charge density (and a larger effective saturation scale of the target; note that the $Q_s$ in this ratio is the parameter from IPSat, not the effective saturation scale that one would extract).

\begin{figure*}[tb]
  \centering
  \subfloat[$\gamma+p\to\jpsim+p$ cross section as a function of center-of-mass energy compared to ALICE~\cite{ALICE:2014eof,ALICE:2018oyo}, H1~\cite{H1:2005dtp,H1:2013okq}, ZEUS~\cite{ZEUS:2002wfj}, and LHCb~\cite{LHCb:2018rcm,LHCb:2024pcz} data. Datapoints with open markers are not included in the fit.]{%
    \includegraphics[width=0.48\textwidth]{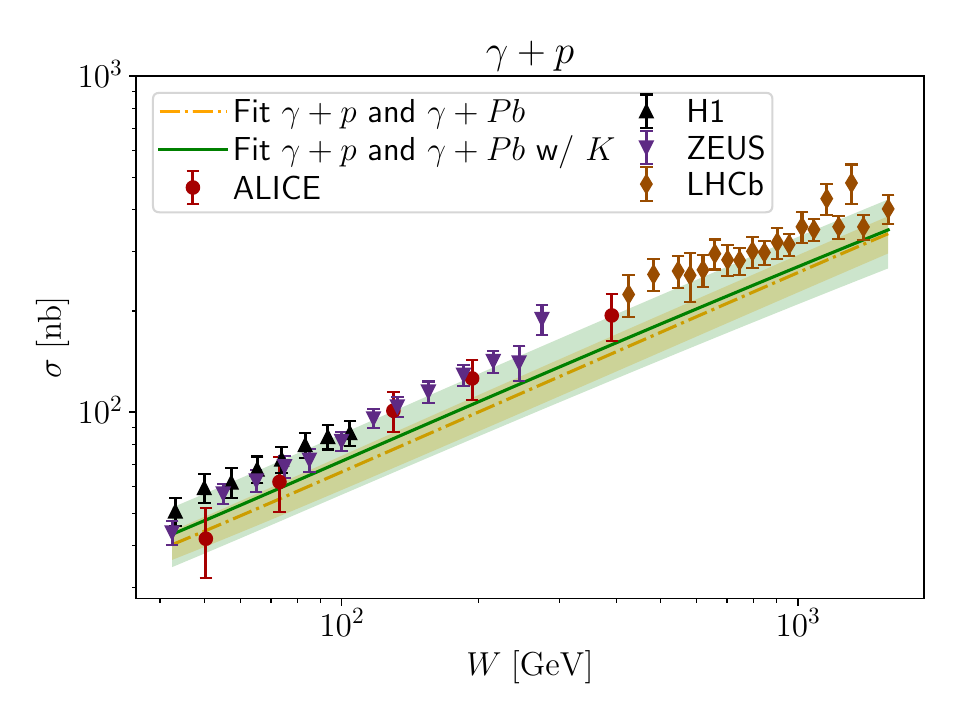}%
    \label{fig:gammap_with_without_K}%
  }
  \hfill
  \subfloat[$\gamma+\mathrm{Pb}\to\jpsim+\mathrm{Pb}$ cross section as a function of center-of-mass energy compared to ALICE~\cite{ALICE:2023jgu} and CMS~\cite{CMS:2023snh} data.]{%
    \includegraphics[width=0.48\textwidth]{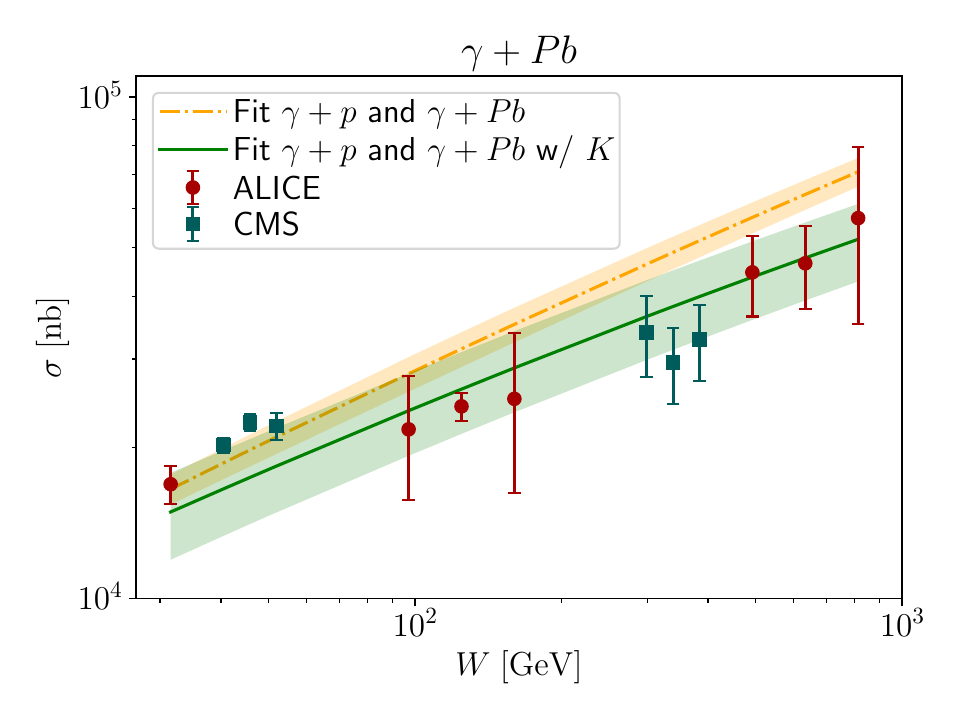}%
    \label{fig:gammaPb_with_without_K}%
  }
  \hfill
  \subfloat[Coherent (smaller $|t|$) and incoherent (larger $|t|$) \jpsi spectra in $\gamma+p$ compared to H1 data~\cite{H1:2013okq}.]{%
    \includegraphics[width=0.48\textwidth]{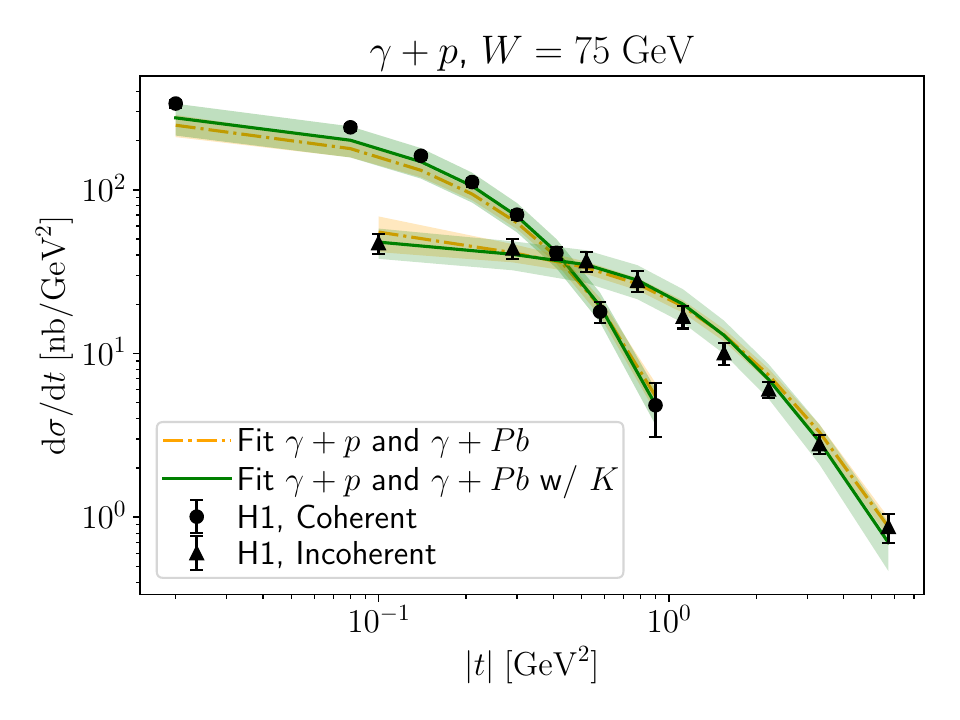}%
    \label{fig:gammap_t}%
    
  }
  \hfill
  \subfloat[Coherent (smaller $|t|$) and incoherent (larger $|t|$) \jpsi spectra in $\gamma+\mathrm{Pb}$ compared to ALICE data~\cite{ALICE:2021tyx,ALICE:2023gcs}.]{%
    \includegraphics[width=0.48\textwidth]{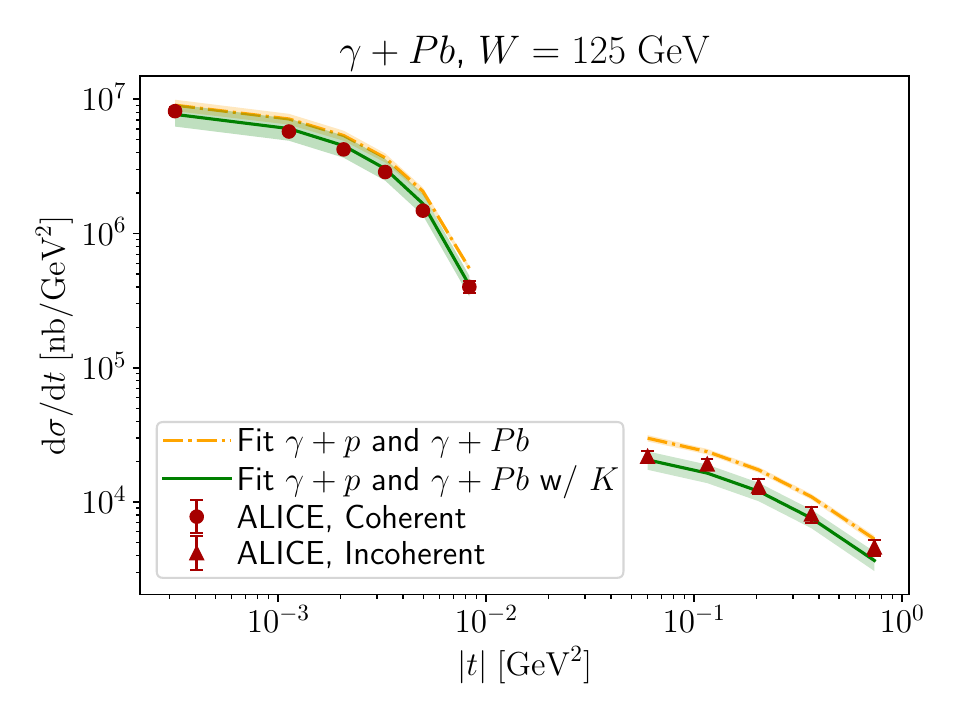}%
    \label{fig:gammaPb_t}%
  }
  \caption{Integrated $W$ dependent and $|t|$-differential cross sections from the fit containing $\gamma+p$ and $\gamma+\mathrm{Pb}$ data in the standard parameter setup (dash-dotted) and including an additional $K$ factor (full) the uncertainty bands indicate the 68\% credible intervals from 25 posterior sample runs of the model.}
  \label{fig:combined_combinedfits}
\end{figure*}

Both the setup with $\Kwf\equiv 1$ and the extended setup with $\Kwf$ as a free parameter are compared to the available energy-dependent $\gamma+p\to\jpsim+p$ and $\gamma+\mathrm{Pb}\to\jpsim+\mathrm{Pb}$ data in Figs.~\ref{fig:gammap_with_without_K} and~\ref{fig:gammaPb_with_without_K}. 
The setup with $\Kwf\equiv 1$ is not compatible with either dataset, underestimating most points in the $\gamma+p$ cross section and overestimating the $\gamma+\mathrm{Pb}$ cross section for $W>100\;\gev$. 
The larger saturation scale in the extended setup now yields stronger nuclear suppression, particularly at high center-of-mass energies, which also leads to a slower energy dependence in the cross section. 
In this case, a good description of all available data is obtained as illustrated by the relatively good $\chi^2$ value quoted in Table~\ref{tab:parameters}.

The setup with $\Kwf\neq 1$ is also clearly preferred by the coherent $t$ spectra shown in Figs.~\ref{fig:gammap_t} and~\ref{fig:gammaPb_t}. 
As discussed in Ref.~\cite{Mantysaari:2022sux}, the coherent spectrum in $\gamma+\mathrm{Pb}$ is sensitive to saturation effects. 
In particular, with the larger saturation scale obtained in the $\Kwf\neq 1$ setup, an excellent description of this data is achieved. 
A good description of the incoherent cross section at all $t$ is also obtained in the extended setup.

To conclude, it is possible to simultaneously describe the available \jpsi photoproduction data with both proton and nuclear targets using a single unified CGC-based framework. 
This agreement, however, requires a relatively small $\Kwf\sim 0.3$ and consequently relatively large proton and nuclear saturation scales. 
It has been previously shown in Ref.~\cite{Mantysaari:2018zdd} that if one constrains the model parameters by using the exclusive cross section data, the inclusive structure function data is typically overestimated. 
In the extended model the proton saturation scale is larger, and consequently this tension between the inclusive and exclusive datasets would be even larger.\footnote{If the $K$ factor is interpreted as representing an uncertainty in the \jpsi wave function, it should not be applied to the inclusive cross section.}

Similarly to Sec.~\ref{sec:standardsetup}, we have also performed a Bayesian analysis including only the $\gamma+\mathrm{Pb}$ data, but using the extended model with $\Kwf$ as a free parameter. 
This analysis suggests a preference for $K\sim 0.4$, with a tail extending to larger values of $K$.
This reaffirms that the requirement to introduce $\Kwf\neq 1$ originates from the $\gamma+\mathrm{Pb}$ data. 
In the future, we plan to investigate whether these conclusions will change when the recent ATLAS measurement (not yet published at the time of writing)~\cite{ATLAS:2025uxr} is included in the analysis.

\section{Conclusions}
\label{sec:conclusions}
Motivated by the previously observed difficulty of CGC-based calculations to simultaneously describe diffractive vector meson production in $\gamma+p$ and $\gamma+\mathrm{Pb}$ collisions, we performed a global Bayesian analysis, including experimental data from both systems. This allowed us to determine if there is any parameter set that allows for a simultaneous description of both systems. 
The description of the nucleon structure at the initial $\xpom=0.01$ is based on an extended (by impact parameter dependence) McLerran-Venugopalan model, there are no free parameters when moving from proton to nuclear targets, and the energy dependence is obtained by solving the JIMWLK equation.
After confirming that the model with parameters constrained by only either $\gamma+p$ or $\gamma+\mathrm{Pb}$ data cannot describe the other system well, we performed a global analysis with the default setup, which includes 7 parameters and has been used in previous studies, and an extended model, which includes a free parameter $K$ that scales all cross sections equally. 

A good fit to all data could only be achieved in the extended model, while the best fit within the default model underestimated $\gamma+p$ data and overestimated $\gamma+\mathrm{Pb}$ data at large center of mass energies $W$. 
The preferred value of the scaling factor $K$ was found to be close to 0.3, indicating a large modification of the default cross-section normalization. 
One could interpret the need for this factor by arguing that uncertainties in the vector meson wave function and/or next-to-leading order corrections are large and can modify the cross sections significantly. 

The reason for the improvement observed when including a $K$-factor of approximately 0.3 is easily understood. 
The resulting reduction in cross section is compensated by a preference for larger color charge densities, increasing the effective saturation scale $Q_s$ in both protons and nuclei. 
Because of the non-linear dynamics at work, this leads to stronger nuclear suppression and hence slower energy evolution in nuclei compared to protons, which is preferred by the experimental data. 

This indicates that calculations performed entirely in the linear regime should not be able to describe the $\gamma+p$ and $\gamma+{\rm Pb}$ data simultaneously, as was also shown in a comparison between calculations using the non-linear Balitsky–Kovchegov (BK) and linear Balitsky–Fadin–Kuraev–Lipatov (BFKL) equations~\cite{Lipatov:1976zz,Kuraev:1977fs,Balitsky:1978ic} in~\cite{Penttala:2024hvp}. NLO BFKL calculations find an improvement, but tension with the $\gamma+{\rm Pb}$ data remains~\cite{Hentschinski:2025ovo} (also see~\cite{Jones:2016icr} for a related study in $\gamma+p$). A future Bayesian analysis utilizing BFKL instead of JIMWLK evolution will help quantify the need for non-linear effects.
Calculations of exclusive vector meson production in $\gamma+{\rm Pb}$ collisions using nuclear parton distribution functions (nPDFs)~\cite{Eskola:2022vaf} within the collinear factorization framework~\cite{Collins:1989gx} show good agreement with experimental data, yet scale and nPDF uncertainties are large. 
Resumming higher-order corrections can significantly suppress the large scale uncertainty, which has been done for $\gamma+p$ collisions in~\cite{Flett:2024htj} using high-energy factorization~\cite{Catani:1990xk,Catani:1990eg,
Collins:1991ty,Catani:1994sq}, but to our knowledge is still missing for the $\gamma+$Pb case. 



We explored further extensions of the model, including parameters that allow for deviations of the nucleon shape from a Gaussian, allow for a different number of hot spots, or modify the dipole amplitude at small dipole sizes. 
Based on a study of the Bayes factors that compare the extended models to the default one, neither of these modifications was found to significantly improve the fit, which is in contrast to the model including the scaling factor $K$. 

This study motivates work towards improving the theoretical description of vector meson production within the CGC. 
This includes extensions to next-to-leading order~\cite{Mantysaari:2022kdm}, improved descriptions of the non-perturbative physics~\cite{Lappi:2020ufv}, including the hadronization into the vector meson, and improvements of the initial state descriptions beyond the MV-model, especially for the proton~\cite{Dumitru:2020gla,Dumitru:2023sjd}. 

For further exploration of saturation effects, it would be beneficial to collect data from other collision systems like $\gamma+\mathrm{O}$, $\gamma+\mathrm{Ne}$, $\gamma+\mathrm{Xe}$, $\gamma+\mathrm{Nd}$, or $\gamma+\mathrm{U}$~\cite{Jia:2022ozr}.

\section*{Data availability statement}
We provide the following numerical results of this work for community use:
\begin{itemize}
    \item A snapshot of the of the \texttt{IPGlasmaFramework} code to produce model predictions~\cite{IPGlasmaFramework}, container file available in Ref.~\cite{mantysaari_2025_15880667}
    \item Training datasets specified in Sec.~\ref{sec:results}~\cite{mantysaari_2025_15880667}
    \item Bayesian inference Python scripts~\cite{hendrik_roch_2025_15879411}
    \item Trained model emulators ~\cite{mantysaari_2025_15880667}
    \item Streamlit application using trained emulators to predict cross sections at adjustable model parameters~\cite{streamlit}
    \item Full posterior distributions and posterior samples used in this work~\cite{mantysaari_2025_15880667}
    \item Posterior sample results from the full model runs~\cite{mantysaari_2025_15880667}
\end{itemize}

\begin{acknowledgements}
We thank G.~Contreras for providing the correlated uncertainties for the ALICE $\gamma+\mathrm{Pb}\to\jpsim + \mathrm{Pb}$ data.
H.M. is supported by the Research Council of Finland, the Centre of Excellence in Quark Matter, and projects 338263 and 359902, and under the European Research Council (ERC, grant agreements No. ERC-2023-101123801 GlueSatLight and No. ERC-2018-ADG-835105 YoctoLHC).
This work is supported by the U.S. Department of Energy, Office of Science, Office of Nuclear Physics, under DOE Contract No.~DE-SC0012704 (B.P.S.), DOE Award No. DE-SC0021969 (C.S.) and DE-SC0024232 (C.S. \& H.R.), and within the framework of the Saturated Glue (SURGE) Topical Theory Collaboration (F.S., B.P.S., W.Z.).
H.R. and W.Z. were supported in part by the National Science Foundation (NSF) within the framework of the JETSCAPE collaboration (OAC-2004571).
C.S. acknowledges a DOE Office of Science Early Career Award.
This research was done using resources provided by the Open Science Grid (OSG)~\cite{Pordes:2007zzb,Sfiligoi:2009cct,OSPool,OSDF}, which is supported by the National Science Foundation awards \#2030508 and \#2323298. 
F.S. is supported by the Laboratory Directed Research and Development of Brookhaven National Laboratory and RIKEN-BNL Research Center. 
Part of this work was conducted while F.S. was supported by the Institute for Nuclear Theory of the U.S. DOE under Grant No. DE-FG02-00ER41132. 
Part of the numerical simulations presented in this work were performed at the Wayne State Grid, and we gratefully acknowledge their support.
The content of this article does not reflect the official opinion of the European Union, and responsibility for the information and views expressed therein lies entirely with the authors.
\end{acknowledgements}

\bibliographystyle{JHEP-2modlong}
\bibliography{refs, non-inspire}

\appendix

\end{document}